\documentclass[prl,twocolumn,groupedaddress,nofootinbib, superscriptaddress]{revtex4-1}
\pdfoutput=1

\usepackage{amsmath,amssymb,amscd}         
\usepackage{listings}
\usepackage{dsfont}
\usepackage{slashed}
\usepackage{soul}
\usepackage[normalem]{ulem}
\usepackage[dvipsnames]{xcolor}

\usepackage{hyperref}
\hypersetup{
    colorlinks=true,
    citecolor=blue,
    linkcolor=blue,
    urlcolor=magenta,
}
\usepackage[capitalise]{cleveref}
\usepackage[pdftex]{graphicx}
\usepackage{epstopdf}
\usepackage{subfigure}
\usepackage{epsfig}
\begin{document}

\rightline{\tiny KCL-PH-TH/2023-44}

\title{Positivity and the Electroweak Hierarchy}


\author{Joe Davighi}
\affiliation{Physik-Institut, Universit\"at Z\"urich, CH-8057 Z\"urich, Switzerland}

\author{Scott Melville}
\affiliation{Astronomy Unit, Queen Mary University of London, Mile End Road, London, E1 4NS, U.K.}

\author{Ken Mimasu}
\author{Tevong~You}
\affiliation{Theoretical Particle Physics and Cosmology Group, Department of Physics, King’s College London, London WC2R 2LS, UK}

\date{August 2023}

\begin{abstract}
\noindent We point out that an unnatural hierarchy between certain higher-dimensional operator coefficients in a low-energy Effective Field Theory (EFT) would automatically imply that the Higgs' vacuum expectation value is hierarchically smaller than the EFT cut-off, assuming the EFT emerged from a unitary, causal and local UV completion. Future colliders may have the sensitivity to infer such a pattern of coefficients for a little hierarchy with an EFT cut-off up to $\mathcal{O}(10)$ TeV. 
\end{abstract}

\maketitle 


\section{Introduction} 

\noindent The electroweak hierarchy problem is made all the more puzzling by the discovery of a Higgs boson with no signs of accompanying new physics beyond the Standard Model to solve its infamous naturalness problem; scalar masses unprotected by any symmetry would naturally lie at the Effective Field Theory (EFT) cut-off scale unless fine-tuning occurs in the UV theory. Positivity bounds are powerful connections between certain EFT coefficients and basic properties of the underlying UV theory (see \cite{deRham:2022hpx} and references therein), and here we explore a potential connection between positivity and the electroweak hierarchy.

The simplest positivity bounds carve the space of EFT coefficients into two regions: (1) values that satisfy the bounds and hence could have arisen from a unitary, causal and local UV completion; and (2) values that violate the bounds and have no such UV completion. Conventional efforts to solve the hierarchy problem live in Region 1. They typically aim to reconcile the electroweak hierarchy with EFT expectations by extending the symmetries of the Standard Model (SM). However, the absence of the necessary new physics at the weak scale has motivated looking for more ``exotic" QFTs to address the hierarchy problem {\it e.g.} Refs.~\cite{Craig:2022uua, Lust:2017wrl, Craig:2019zbn}. They aim to break the rules of EFT by some as yet unknown UV/IR mixing, which is expected to violate decoupling and locality. Such exotic QFTs may live in either Regions 1 or 2. If measurements of EFT coefficients place us in Region 2, this would be a smoking gun for an exotic UV theory. 

\begin{figure}[t!]
    \centering
    \includegraphics[width=0.8\linewidth]{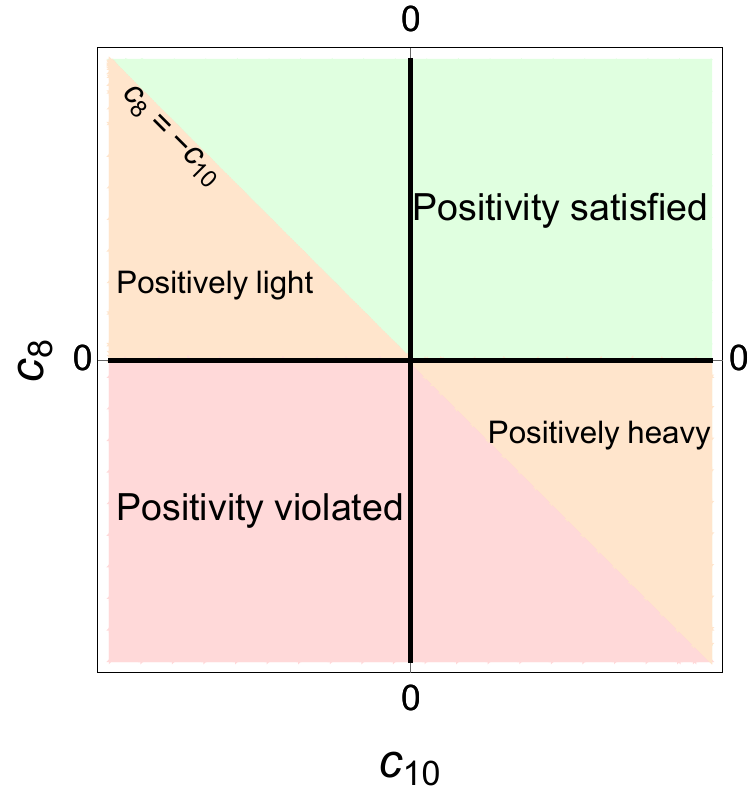}
    \caption{Possible values for a particular dimension-8 (dimension-10) operator coefficient $c_8$ ($c_{10}$). The subset of Region 1 in which positivity bounds are satisfied for any vev is shown in green. The remainder of Region 1, in which positivity is only satisfied for a restricted range of vevs (Eqs.~\ref{eq:mainvevbound} or~\ref{eq:mainvevbound2}) is coloured orange---we refer to this as Region $\bar{1}$. Region 2, in which the positivity bounds are always violated, is shown in red.
    }
    \label{fig:EFTspace} 
\end{figure}

We point out that there is a subset of Region 1 in which positivity is only satisfied for a restricted range of the Higgs vacuum expectation value (vev), $v$, relative to the EFT cut-off scale, $\Lambda$, that we denote by Region $\bar{1}$. In particular, if there exist unitary, local, and causal UV theories that map to a specific pattern of dimension-8 and dimension-10 operator coefficients $c_8$ and $c_{10}$, respectively, with $|c_{10}| \gg |c_8|$ and appropriate signs, then by positivity they necessarily also have a hierarchy that satisfies
\begin{equation}
    \frac{v^2}{\Lambda^2} < \frac{|c_8|}{|c_{10}|}  \qquad \text{(Positively light)}\, .
    \label{eq:mainvevbound}
\end{equation}
We define a Higgs whose vev satisfies Eq.~\ref{eq:mainvevbound} living in the subset of Region $\bar{1}$ where $c_8 > 0$ and $c_{10} < 0$ as being ``positively light". 
Another subset of Region $\bar{1}$ with the signs reversed requires instead a Higgs satisfying a lower bound on its vev, that we call ``positively heavy", 
\begin{equation}
    \frac{v^2}{\Lambda^2} > \frac{|c_8|}{|c_{10}|}  \qquad \text{(Positively heavy)}\, .
    \label{eq:mainvevbound2}
\end{equation}
The EFT coefficient space is illustrated in Fig.~\ref{fig:EFTspace}. 
Analogous positivity bounds involving dimension-6 operator coefficients can also be derived under additional UV assumptions.

Often, varying the parameters of a given UV theory can only produce a particular range of EFT coefficients at low energies. Our novel use of positivity shows that in any unitary, causal and local UV theory, once its parameters have been partially fixed so that $|c_8| \ll |c_{10}|$ (with $c_8 > 0$ and $c_{10} < 0$), then no matter how the remaining parameters are varied the Higgs vev may only take the unnaturally small values in Eq.~\ref{eq:mainvevbound}. Since these UV theories produce a characteristic hierarchy between particular higher-dimension operators in the low-energy EFT, this scenario could be identified in future experiments.

A positively light Higgs would not, {\it per se}, be a solution to the hierarchy problem; as we will see, Region $\bar{1}$ from an EFT perspective is a fine-tuning in the ratio of higher-dimensional operator coefficients. Nevertheless, it is interesting that a correlated tuning of {\it a priori} unrelated quantities in the EFT --- the dimension $\leq 4$ scalar potential and higher-dimensional operator coefficients --- could be a sign of some UV mechanism restricting the allowed spectrum of Higgs vevs. A potential measurement establishing us to be in Region $\bar{1}$ would strongly motivate looking for such a UV-completion.

\section{Positivity without vevs}
\label{sec:positivity}

\noindent We begin by reviewing the known positivity bounds for EFTs in which every field has already been expanded around a unique stable vev. In particular, we summarise the two-sided bounds that forbid unnatural hierarchies between successively higher-dimensional operator coefficients (that are not in conflict with the positivity bounds with vevs that we discuss below), 
and recall the assumptions under which one may derive dimension-6 positivity bounds.

The central object that bridges between the EFT and the UV is the scattering amplitude for the elastic process $AB \to AB$. 
This can be computed using the low-energy EFT
at centre-of-mass energies $s < \Lambda^2$, where $\Lambda$ is the EFT cut-off.
In the complex $s$-plane, we therefore have an accurate determination of the amplitude inside a disc of radius $\Lambda^2$. 
Subtracting any singularities which appear inside this disc\footnote{
It will be convenient to also subtract any $t$-channel poles appearing in the EFT amplitude. Note that we are working with a non-gravitational EFT, i.e. in the decoupling limit $M_P \to \infty$ where the $t$-channel pole from graviton exchange can be neglected. Otherwise the subtraction of this $s^2/t$ term would lead to an object that violates the Froissart bound, and for which the residue at $s= \infty$ in Eq.~\ref{eqn:dispersion} only vanishes for $i \geq 3$ \cite{Adams:2006sv, Bellazzini:2015cra}. } leads to an analytic amplitude which can be written as,
\begin{align}
 \mathcal{A} (s,t) = \sum_{i,j = 0} \, a_{i,j} \, \frac{s^i \, t^j }{ \Lambda^{2i + 2j} }  \; , 
 \label{eqn:A_to_a}
\end{align}
where $t$ is the usual momentum transfer and increasing $i+j$ corresponds to increasing orders in the EFT derivative expansion. 
The EFT coefficients $a_{i,j}$ can be extracted from Eq.~\ref{eqn:A_to_a} by $\partial_s^i \partial_t^j \mathcal{A} (s,t) |_{s=t=0}$, or equivalently by integrating $\partial_t^j \mathcal{A}/s^{i+1}$ around a closed contour which encircles the origin in the complex $s$-plane.
The assumption of causality (or, more precisely, analyticity) in the UV implies that this low-energy contour can be deformed into a high-energy contour that encircles the singularities of the UV theory \cite{Adams:2006sv, deRham:2017avq},
\begin{align}
 a_{i,j} = \frac{ \partial_t^j }{j!} \Bigg[&  \int_{\Lambda^2}^{\infty} \frac{ds}{2 \pi i} \left( \frac{ \text{Disc}_s \mathcal{A} (s,t)  }{s^{i+1}} - \frac{ \text{Disc}_u \mathcal{A} (u,t)  }{u^{i+1}}  \right)  \nonumber \\
 &- \text{Res}_{s=\infty} \left( \frac{\mathcal{A} (s,t) }{ s^i } \right) \Bigg] \; , 
 \label{eqn:dispersion}
\end{align}
where $u = -s -t + 2 m_A^2 + 2m_B^2$ is the third Mandelstam variable, $\text{Disc}_z$ is the discontinuity across the real $z$-axis, and $\text{Res}_{s=\infty}$ is the residue at infinity. This ``sum rule'' explicitly connects the EFT coefficients to the underlying UV physics.

The second assumption we make of the UV is that time evolution is unitary. In terms of the amplitude, this implies $\text{Disc}_s \, \mathcal{A} \geq 0$ via the well-known optical theorem~\footnote{
While there are different conventions in use for the overall phase of the amplitude, this can always be chosen so that the optical theorem implies a positive discontinuity.
}. If the quantum numbers of $A$ and $B$ are chosen so that there is a trivial crossing relation between the $s$- and $u$-channel, then unitarity also implies that $\text{Disc}_u \, \mathcal{A} \geq 0$ \footnote{
Positivity bounds can also be derived more generally with a non-trivial crossing relation between $s$- and $u$-channel \cite{Bellazzini:2016xrt, deRham:2017zjm, deRham:2018qqo, Melville:2019tdc, Davighi:2021osh}.
}. 
Our third UV assumption, locality (in the form of the Froissart bound), guarantees that the residue at infinity vanishes for all $i \geq 2$. 
Consequently, the $a_{2n,0}$ coefficients in the EFT expansion must obey the bound \cite{Adams:2006sv, deRham:2017avq, Bellazzini:2017fep, deRham:2017xox},
\begin{align}
 a_{2n, 0 } \geq 0 \;  ,
 \label{eqn:pos1}
\end{align}
for all $n \geq 1$ if its UV completion is to be causal, unitary and local~\footnote{
We assume Lorentz invariance throughout, and also the existence of an $S$-matrix at high-energies. 
}. This bound can only be saturated for any particular $a_{2n,0}$ if every EFT coefficient vanishes and the theory is trivially free at all energies \cite{deRham:2017imi}.

Analogous bounds can also be derived for the $a_{2n,1}$ coefficients (again with $n\geq 1$), and these take the form,
\begin{align}
 - \beta_n \;  a_{2n,0} \leq  a_{2n,1}  \leq + \alpha_n \; a_{2n,0}  \; .
 \label{eqn:pos2}
\end{align}
The positive parameters $\alpha_n$ and $\beta_n$ depend on which scattering process is considered, but ultimately there is always a two-sided bound which forbids arbitrarily small tunings of $a_{2n,0}$ relative to $a_{2n,1}$.
Concretely, $2 \beta_n = 2n+1$ for scalar fields \cite{deRham:2017avq} and $2 \beta_n = 2 n + 1 - |h_A  + h_B | - | h_A - h_B |$ for massless spinning fields with helicities $h_A$ and $h_B$ \cite{Davighi:2021osh}~\footnote{
For massive spinning fields, $2 \beta_n = 2 n + 1 - |h_A  + h_B | - | h_A - h_B |_{\rm min}$, where the minimum is over $h_{A,B}$ at fixed total $h_A + h_B$.
}.
For a general $\mathcal{A}_{AB \to AB}$ amplitude,
$ \alpha_1 / (16 \pi^2) = 5! / a_{2,1}$ \cite{Bellazzini:2017fep, Bellazzini:2020cot} and the upper bound in Eq.~\ref{eqn:pos2} roughly corresponds to $a_{2,0}$ being one loop factor away from $a_{2,1}^2$. For a maximally crossing symmetric amplitude like $\mathcal{A}_{AA \to AA}$ for a real scalar $A$, the additional constraints from crossing can be used to derive further bounds \cite{Tolley:2020gtv}, the strongest of which in four spacetime dimensions is $\alpha_1 \approx 5.3$ \cite{Caron-Huot:2020cmc}. For such amplitudes, there can be no hierarchy between $a_{2,1}$ and $a_{2,0}$. 

The positivity bounds of Eqs.~\ref{eqn:pos1} and \ref{eqn:pos2} are naturally formulated in terms of the coefficients appearing in the amplitude.
In practice, it is often useful to apply these bounds to the EFT coefficients appearing in a Lagrangian.
For instance, consider, 
\begin{equation}
    \mathcal{L}_{\rm EFT} = \sum_{d=0}^{\infty} \;  \bar{c}_d \;  \frac{ \mathcal{O}_d }{\Lambda^{d-4}} \, ,
\end{equation}
where $\mathcal{O}_d$ denotes a generic dimension-$d$ operator, $\bar{c}_d$ is its constant coefficient, and there is an implicit sum over all operators at each dimension. 
At tree-level, it is the dimension-8 operators which contribute to the $s^2$ part of the amplitude and hence $a_{2,0}$ is a linear combination of the $\bar{c}_8$ coefficients\footnote{
Depending on the EFT operator basis, there may also be a contribution to $a_{2,0}$ from pairs of dimension-6 operators.
}. 
The bound Eq.~\ref{eqn:pos1} then implies that some $\bar{c}_8$ (or a linear combination of them) must be sign-definite if a unitary, causal and local UV completion is to exist.
Similarly, $a_{2,1}$ is a linear combination of the $\bar{c}_{10}$ coefficients and Eq.~\ref{eqn:pos2} implies that these $\bar{c}_{10}$ cannot be tuned arbitrarily large relative to $\bar{c}_8$. As a concrete example, $\bar{c}_8 \mathcal{O}_8 = \bar{c}_8 (\partial \phi )^4$ for a real scalar field $\phi$ must have $\bar{c}_8 \geq 0$, and the higher-derivative correction $\bar{c}_{10} \mathcal{O}_{10} = \bar{c}_{10} ( \partial \phi )^2 (\partial \partial \phi)^2$ may not be hierarchically larger than $\bar{c}_8$.
Note that while field redefinitions, as well as the freedom to redefine the operator basis of $\mathcal{O}_d$, can be used to shuffle the $\bar{c}_d$ coefficients around, it is the amplitude coefficients $a_{i,j}$ which are the invariant (observable) quantities bounded by positivity---given an EFT operator, one must always check how it contributes to the $a_{i,j}$ before declaring a positivity bound on its coefficient.

Finally, it was recently noticed in Ref.~\cite{Davighi:2021osh} that for sufficiently large helicities the parameter $\beta_n$ can vanish, and as a result $a_{2n,1} \geq 0$ obeys a simple positivity bound. 
In particular, for the scattering of two spin-1/2 fields with helicities $h_A = h_B = +1/2$, the amplitude must obey,
\begin{align}
 a_{0,1} \geq 0 \; , 
\end{align}
in order to have a UV completion which is causal, unitary and converges at large $s$ slightly faster than the Froissart bound (so that the residue at infinity vanishes for $a_{0,1}$). This subset of local UV completions contains, for instance, all tree-level completions with no $t$-channel exchange \cite{Adams:2008hp,Remmen:2020uze, Remmen:2022orj}. 
Since $a_{0,1}$ is a linear combination of the $\bar{c}_6$ coefficients, such a UV completion can only exist if some dimension-6 operator coefficients are sign-definite.

\section{Positivity with vevs}

\noindent Now we turn to the main focus of this work, which is the effect on this story of a scalar field having a non-trivial vev.
To illustrate this most simply, consider an EFT which contains a dimension-8 and dimension-10 operator of the form,
\begin{align}
 \mathcal{L}_{\rm EFT} [ H ] =   c_8 \frac{\mathcal{O}_8}{\Lambda^4}  +   c_{10}  \frac{ |H|^2 \mathcal{O}_8 }{\Lambda^6} \, ,
 \label{eqn:EFT2}
\end{align}
where $\mathcal{O}_8$ is the only interaction that could contribute to the $s^2$ part of an $AB \to AB$ scattering amplitude, and $H$ is an additional (possibly complex) scalar field. We will often identify $H$ with the complex Higgs doublet of the SM,\footnote{In the context of the electroweak theory, our discussion applies only to UV theories that match onto the SMEFT at low energy. 
We do not consider more general UV theories that can be matched at low energies only onto a non-linearly realised electroweak chiral Lagrangian with a singlet Higgs, known as the `Higgs EFT (HEFT)' (see {\it e.g.} Ref.~\cite{Cohen:2020xca} and references therein).} although the discussion here is more general. 
We assume the potential for $H$ has a stable vev at $|H| = v$. 
We also assume that a well-defined $S$-matrix element for $AB \to AB$ scattering exists on this background.

Focussing on low-energy perturbations about this vacuum, i.e. integrating out $H$, produces the simpler EFT,
\begin{align}
 \mathcal{L}_{\rm EFT} [ v ] =   \bar{c}_8 \frac{ \mathcal{O}_8 }{\Lambda^4}  \, ,
 \label{eqn:EFT3}
\end{align}
where
\begin{equation}
    \bar{c}_8 = c_8 + \frac{v^2}{\Lambda^2} c_{10} \, .
    \label{eq:cbar8}
\end{equation}
Provided we have normalised $\mathcal{O}_8$ so that $a_{2,0} = \bar{c}_8$, the positivity bound of Eq. \ref{eqn:pos1} then requires,
\begin{align}
 c_8 + \frac{v^2}{\Lambda^2} c_{10} \geq 0 \; . 
\label{eqn:pos3}
\end{align}
The underlying UV theory may only be unitary, causal, local \emph{and contain $v$ as a stable vev} if this bound is satisfied\footnote{
If there is more than one stable vacuum, then there will be one bound of the form Eq.~\ref{eqn:pos3} for each vev. Some regions of the EFT parameter space may only admit UV completions (satisfy the positivity bounds) around some of these vevs \cite{Melville:2022ykg}, which corresponds to the fact that distant minima in the EFT potential can be de-stabilised when integrating in heavy states.
}. 
The implication of positivity bounds around different vevs was recently studied in Ref.~\cite{Melville:2022ykg} in the context of cosmological EFTs, where demanding the existence of different vacua ({\em e.g.} flat versus expanding spacetime \cite{Melville:2019wyy, deRham:2021fpu}) places different constraints on the low-energy coefficients. 
Here, we turn that logic around and point out that for certain values of the coefficients, the vev is effectively constrained by the requirements of unitarity, causality and locality in the UV, as in Eq.~\ref{eq:mainvevbound}.
In particular, when there is a hierarchy $|c_{8}| \ll |c_{10}|$, a negative value of $c_{10}$ (with $c_8 > 0$) can only satisfy this bound providing $v$ is hierarchically smaller than $\Lambda$. 
Consequently, any standard UV theory that produces a low-energy EFT with $|c_8|/|c_{10}| \ll 1$ with $c_8 c_{10} < 0$ could only ever produce a restricted range of Higgs vevs.

The new insight here is that the low-energy coefficient $\bar{c}_8$ in $\mathcal{L}_{\rm EFT} [v]$ 
receives contributions from coefficients of higher-dimension operators once
we partially UV complete the theory by introducing the radial modes of $H$. 
More generally than Eq.~\ref{eqn:EFT2}, this partial UV completion takes the form, 
\begin{align}
 \mathcal{L}_{\rm EFT} [ H ] =  f_8 (\xi)  \frac{\mathcal{O}_8}{\Lambda^4}, \qquad \xi :=\frac{|H|^2}{\Lambda^2}   \; .  \label{eqn:EFT4}
\end{align}
Positivity of low-energy $AB \to AB$ scattering around the $|H|=v$ vacuum then requires, 
\begin{align}
 f_8 (v^2/\Lambda^2 ) \geq 0 \; . 
 \label{eqn:pos4}
\end{align}
This can impose a restriction on the vev $v$ whenever the function $f_8(\xi)$ violates the natural EFT power counting  in which $f_8 (\xi) = \sum_n f_{8,n} \xi^n$ with Taylor coefficients $f_{8,n} \sim \mathcal{O}(1)$.
With this power counting, $f_8 (\xi)$ is expected to be very flat near the vacuum point $\xi \approx v^2/\Lambda^2$, since all derivatives would be suppressed by powers of $v^2/\Lambda^2 \ll 1$. 
Any violation of this power counting, in which one or more derivatives of $f_8$ become large and with appropriate signs, would produce a bound on $v^2/\Lambda^2$ in terms of $f_8$ evaluated at a different value of $\xi$ (away from the vacuum point). 
For instance, when $f_8  (0)$ is unnaturally small the positivity bound Eq.~\ref{eqn:pos4} can be written as,  
\begin{align}
     \frac{v^2}{\Lambda^2} + \mathcal{O} \left( \frac{v^4}{\Lambda^4} \right)  \leq  \frac{ |f_8 ( 0 )| }{ | f_8' (v^2/\Lambda^2 )| } \ll 1  \, ,
     \label{eq:posvevboundfunctionalform}
\end{align}
whenever $f_8 (0) > 0$ and $f'_8 (v^2/\Lambda^2 ) < 0$. 

Furthermore, for UV theories with a super-Froissart convergence we would arrive at the same conclusion for dimension-6 operators of the form, 
\begin{align}
 \mathcal{L}_{\rm EFT} [ H ] =  f_6 \left( \xi \right)  \frac{ \mathcal{O}_6 }{ \Lambda^2 } \; , 
\end{align}
where $\mathcal{O}_6$ contributes 
to $a_{0,1}$ for four-Fermi scattering with aligned helicities.
While the restriction to super-Froissart growth makes such bounds less general, observational prospects are better for these dimension-6 operators (see, for example, Refs.~\cite{Ellis:2020unq, Ethier:2021bye} for current constraints from recent global fits and Ref.~\cite{deBlas:2022ofj} for future sensitivity projections). 

Finally, notice that since $c_{10}$ ($f_8'$) does not contribute to $a_{2,1}$, it is not bounded in relation to $c_8$ ($f_8$) by other positivity bounds like Eq.~\ref{eqn:pos2}. Unlike dimension-10 operators which contain more derivatives, a dimension-10 operator which contains more fields may be tuned much larger than $c_8$---at least, there is no known positivity argument that prevents it. 
Of course, in practice, the ``natural'' expectation is that $c_8 \sim c_{10} \sim \mathcal{O} (1)$ when there are no symmetries or selection rules to suggest otherwise, since this is invariably what happens in the simplest UV completions.  \\

\noindent {\it Example UV completion} --- The unnatural hierarchy in higher-dimensional operator coefficients that a UV theory must have to belong to Region $\bar{1}$ leads us to expect a UV-completion that goes beyond the simplest 
possibilities. In the absence of an explicit realisation of such a UV theory, it is nonetheless useful to see a concrete example of a simple model in which the obstacles can be made explicit.  

Consider a toy model with a heavy real scalar $\phi$ of mass $M$, a light complex scalar $H$ and a Dirac fermion $\Psi$. Their UV interactions include
\begin{equation}
    \mathcal{L}_{UV} \supset -\left( y\phi\bar{\Psi}\Psi + \text{h.c.} \right) - \mu g_3 \phi^3 - \mu g_1 \phi|H|^2 \, ,
\end{equation}
where $y, g_1, g_3$ are dimensionless couplings and $\mu$ is a dimensionful scale. Matching to the EFT Lagrangian Eq.~\ref{eqn:EFT2}, with 
\begin{equation}
    \mathcal{O}_8 = -\bar{\Psi}\Psi\partial_\mu\bar{\Psi}\partial^\mu\Psi \, ,
\end{equation}
using {\tt Matchete}~\cite{Fuentes-Martin:2022jrf} yields
\begin{align}
    c_8 &= \left(y + \bar{y}\right)^2 \, , \\
    c_{10} &= \frac{4 g_1 (3g_3 - g_1)\mu^2}{M^2}\left(y + \bar{y}\right)^2  \, .
\end{align}
We see that $c_8$ has a definite positive sign, while $c_{10}$ can have arbitrary sign depending on our choice for the couplings $g_1, g_3$. Substituting into Eq.~\ref{eqn:pos3}, assuming $c_{10}$ is negative, then gives an apparent bound on the vev, 
\begin{equation}
    \frac{v^2}{M^2} < \frac{|c_8|}{|c_{10}|} = \frac{1}{|g_1(3g_3 - g_1)|}\frac{M^2}{\mu^2} \, .
\end{equation}
However, this cannot impose a hierarchy $v^2 / M^2 \ll 1$ by more than a loop factor, since it would require taking $\mu \gg M$ which is associated with a breakdown of perturbation theory when $(g_i\mu)^2\geq 4\pi M^2$. A more systematic exploration of potential UV completions is warranted. \\

\noindent {\it Experimental prospects} --- We have so far discussed the positivity bound Eq.~\ref{eq:mainvevbound} as a theoretical constraint on the EFT parameter space. To potentially establish this bound experimentally depends on whether a low-energy observer has access to measurements at different energy scales within the EFT. 
From hereon we fix the scalar field $H$ to be the complex Higgs doublet of the electroweak theory. Expanding around the vev, we also define the singlet radial mode $h$ via $H =(0, v+h(x))^T/\sqrt{2}$.

In the $|H|^2=v^2/2$ vacuum of the SM, and at very low energies $E$ below the Higgs mass $m_h$, i.e. $E < m_h < \Lambda$, an experimentalist can only measure the physical combination $\bar{c}_8$ of Eq.~\ref{eq:cbar8}, for example via $AB \to AB$ scattering. One cannot determine the `vev contribution' to $\bar{c}_8$ by doing measurements at these low energy scales. At higher energies, $m_H < E < \Lambda$, but still within the EFT, the Higgs' radial mode $h$ can be produced on-shell and participate in scattering processes, so that $c_{10}$ can be measured {\it e.g.} via the $AB \to ABh$ or $AB \to AB hh$ processes. More precisely, starting from the more general EFT described by Eq.~\ref{eqn:EFT4}, one can relate the function $f_8$ to these scattering processes by expanding the Lagrangian around the vacuum: 
\begin{equation}
    \mathcal{L} = \left[f_8\left(\frac{v^2}{2\Lambda^2}\right)  + f_8^\prime\left(\frac{v^2}{2\Lambda^2}\right) \frac{vh}{\Lambda^2} + \mathcal{O}(h^2)\right] \, \frac{\mathcal{O}_8}{\Lambda^4}\, .
\end{equation}
So, by measuring $AB \to AB$ and $AB \to AB h$ scattering, we can extract the quantities $\frac{1}{\Lambda^4}f_8(v^2/2\Lambda^2) = \frac{\bar{c}_8}{\Lambda^4}$ and $\frac{v}{\Lambda^6}f_8^\prime(v^2/2\Lambda^2)$ respectively. From the former alone, it is possible to ascertain whether the positivity bound Eq.~\ref{eqn:pos4} is satisfied. The latter measurement is of a dimension-9 HEFT operator $h \mathcal{O}_8$, that comes from a dimension-10 SMEFT operator whose coefficient in units of $1/\Lambda^6$ is $c_{10}$. 

However, with these measurements alone one cannot use positivity (say, in the form of Eq.~\ref{eq:posvevboundfunctionalform}) to infer whether the vev is bounded to be hierarchically small. The `vev contribution' to $\bar{c}_8$ is $f_8(v^2/2\Lambda^2)-f_8(0) \approx \frac{v^2}{2\Lambda^2} f_8^\prime(v^2/2\Lambda^2)$ up to a dimension-12 contribution, and we are `sensitive' to the RHS by doing measurements of $AB \to ABh$ at these intermediate energy scales.
But without an independent extraction of the cut-off scale $\Lambda$ we cannot turn this into a bound on $v$. Alternatively, as can be seen from Eq.~\ref{eq:posvevboundfunctionalform}, one could unambiguously establish the bound on $v/\Lambda$ by accessing the value of $f_8$ (or its derivative) at a different value of $|H|^2/\Lambda^2$ ({\em e.g.} zero), say by measuring $AB \to AB(h)$ scattering in a different (meta-)stable vaccum. While this is in principle possible within the EFT, it is difficult to imagine doing any such measurement in practice.

In the absence of a direct determination of $c_8 = f_8(0)$, we can still obtain indirect evidence for being in Region $\bar{1}$ if we were to measure $\bar{c}_8/\Lambda^4$ and $vc_{10}/\Lambda^6$ as above, and thence infer that $\bar{c}_8 \ll |c_{10}|$ given the cut-off scale $\Lambda$ is at least a TeV --- a reasonable assumption given current null results in direct searches~\footnote{If evidence of non-zero dimension-6 EFT coefficients were also to be established in experiment, one could put a more meaningful `prior' on the scale $\Lambda$ that would then imply an even smaller ratio of $\bar{c}_8 / |c_{10}|$ the higher the scale $\Lambda$ is above a TeV.}.
This is expected to be the case in Region $\bar{1}$ since $\bar{c}_8$ is bounded from above by $c_{10}$. 
Therefore, even if we are not able to explicitly establish an upper bound on the Higgs vev through measurements, if we indeed live in Region $\bar{1}$ then a consequence is that we should find an unnatural suppression of a particular ratio of higher-dimensional operator coefficients and no new physics below a certain scale. 

Dimension-8 operators have been constrained at the LHC in various processes~\cite{Bellazzini:2017bkb, Bellazzini:2018paj,Zhang:2018shp,Bi:2019phv, Remmen:2019cyz, Englert:2019zmt, Remmen:2020vts, Bonnefoy:2020yee, Ghosh:2022qqq, Bonnefoy:2020yee, Henriksson:2021ymi, Li:2022rag, Ellis:2023zim, Ellis:2022zdw, Ellis:2021dfa}, with promising prospects for much higher sensitivity at future colliders~\cite{Gu:2020ldn, Fuks:2020ujk, Ellis:2022zdw, Ellis:2021dfa}. While the feasibility of measuring dimension-10 operators are more uncertain, it may be propitious for Region $\bar{1}$ since the dimension-10 effects could then compete with dimension-8. Dimension-10 operators consisting of fields including the two initial-state partons or leptons together with the Higgs and two other particles in the final state could benefit from interference with SM amplitudes and significant energy growth in future high energy colliders. 

As an illustrative example, we consider a set of four-lepton, dimension-8 operators and their dimension-10 counterparts where an $|H|^2$ is attached to the corresponding dimension-8 operator. The dimension-8 operators are defined as 
\begin{align}
    \mathcal{O}_8^{(1)} &= \partial^\nu \left(\bar{e_i}\gamma^\mu e_i\right)\partial_\nu \left(\bar{e}_i\gamma_\mu e_i\right) \, , \notag \\
    \mathcal{O}_8^{(2)} &= \partial^\nu \left(\bar{e}_i\gamma^\mu e_i\right)\partial_\nu\left(\bar{L}_i\gamma_\mu L_i\right) \, , \notag \\
    \mathcal{O}_8^{(3)} &= D^\nu \left(\bar{e}_i L_i\right)D_\nu\left(\bar{L}_i e_i\right) \, , \notag \\
    \mathcal{O}_8^{(4)} &= \partial^\nu \left(\bar{L}_i \gamma^\mu L_i\right) \partial_\nu \left(\bar{L}_i \gamma_\mu L_i\right) \, ,
\end{align}
where $i$ labels the lepton flavour, and $L$ and $e$ denote the left-handed lepton doublets and right-handed (charged) lepton singlets respectively.
The full set of positivity bounds that can be obtained from four-lepton scattering amplitudes was computed in Ref.~\cite{Fuks:2020ujk}.
For example, considering the $e_R e_L\to e_R e_L$ amplitude requires
the coefficient of $\mathcal{O}^{(3)}_8$, namely $c_8^{(3)}$, to be positive. Taking into account the corresponding dimension-10 operator, $\mathcal{O}^{(3)}_{10}=|H|^2\mathcal{O}_8^{(3)}$, the bound now applies to
\begin{align}
    \bar{c}_8^{(3)} = c_8^{(3)} + \frac{v^2}{2\Lambda^2} c_{10}^{(3)}\geq 0.
\end{align}
We note that some of the bounds on this set of operators require going slightly beyond considering individual operators contributing to elastic scattering processes of type $AB\to AB$, as previously discussed. For instance, $\mathcal{O}^{(4)}_8$ is not the only operator that contributes to the elastic amplitudes that can be used to bound its coefficient. Furthermore, the full set of bounds can only be obtained by considering the scattering of arbitrary superpositions of different flavours and chiralities of (anti)leptons \cite{Trott:2020ebl, Zhang:2020jyn}, particularly those involving $\mathcal{O}_8^{(2)}$, which does not contribute to the $s^2$ coefficient of any elastic amplitude involving fixed flavours and chiralities.

\begin{figure}[t!]
    \centering
    \includegraphics[width=1.0\linewidth]{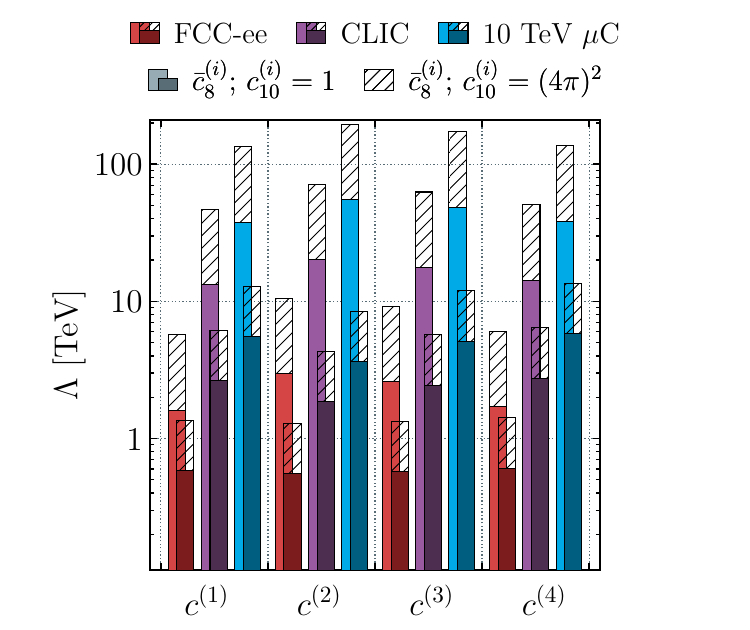}
    \caption{Projected sensitivity of FCC-ee (red), CLIC 3 TeV (purple), and a 10 TeV muon collider (blue) to four-fermion dimension-8 operators (taken from Ref.~\cite{Fuks:2020ujk} for FCC-ee and CLIC) and to their corresponding dimension-10 operators with a $|H|^2$ attached, represented by the lighter and darker shades respectively. The filled bars assume a Wilson coefficient of 1, while the hatched bars assume a value of $(4\pi)^2$. }
    \label{fig:projections}
\end{figure}

In Fig.~\ref{fig:projections}, we show rough projected sensitivities at future lepton colliders to the $c_8^{(i)}$ and $c_{10}^{(i)}$, taking FCC-ee~\cite{FCC:2018byv} and CLIC~\cite{CLIC:2018fvx}, as well as a hypothetical 10 TeV muon collider~\cite{Accettura:2023ked} as typical examples across a range of centre-of-mass energies. Most of the projected sensitivities for the dimension-8 operators are taken from Ref.~\cite{Fuks:2020ujk}, except for the muon collider sensitivities which we computed ourselves.
The relevant processes for bounding the dimension-8 and -10 operators are $e^+e^-\to e^+e^-$ and $e^+e^-\to e^+e^-h$, respectively (replacing $e$ with $\mu$ for the muon collider). In the former case the differential cross-section with respect to the polar angle of the lepton was used as a discriminating variable, and in the latter the differential distribution in the angular separation between the lepton and anti-lepton was used, exploiting the kinematic features of the dominant $e^+e^-\to Zh$ background to define the bin boundaries. We estimated the sensitivity to the dimension-10 operators by a na\"ive statistical error projection based on the number of expected events. For the $e^+e^-$ colliders, the run conditions assumed in Ref.~\cite{Fuks:2020ujk} and summarised in Table~\ref{tab:conditions} were matched as closely as possible, with the exception of the polarised beam at CLIC.
\setlength{\tabcolsep}{5pt}
\renewcommand{\arraystretch}{1.5}
\begin{table}[t!]
  \centering
    \begin{tabular}{lllll}
    \hline
    Collider & \multicolumn{4}{c}{Runs: Energy [GeV] (Luminosity [ab$^{-1}$])} \\
    \hline
    \hline
    FCC-ee\phantom{-} & 161 (10),& 240 (5),& 350 (0.2), &  365 (1.5)\\
    CLIC & 380 (0.5),& 1500 (2), & 3000 (4)&\\
    $\mu$C & 10000 (10) &&&\\
    \hline
    \end{tabular}
  \caption{Run conditions assumed for the projected sensitivities shown in Fig.~\ref{fig:projections}.}
  \label{tab:conditions}
\end{table}

The sensitivity is represented by a scale, $\Lambda$, assuming $\bar{c}^i_8,c^i_{10}=1$ for the dimension-8 (light coloured bars) and dimension-10 (dark coloured bars) coefficients, respectively. The sensitivities indicated by the hatched bars correspond to the maximally strongly coupled scenario of $\bar{c}^{(i)}_8,c^{(i)}_{10}=(4\pi)^2$.
The projected reach on $\Lambda$ increases (decreases) for coefficients induced by strongly (weakly) coupled new physics. All colliders are able to probe energy scales above their centre of mass energies~\footnote{An FCC-hh machine~\cite{FCC:2018byv} could probe even higher energy scales for similar operators involving quarks, but we limit our comparison to lepton colliders here. }. A 3 TeV CLIC or 10 TeV muon collider offer the best prospects for probing the energy frontier associated with the 4-lepton operators we consider, while FCC-ee, even though it is primarily a precision frontier machine, can still reach sensitivity to the 10 TeV scale for dimension-8 operators with large EFT coefficients. For these high-energy lepton colliders, we see that the scale that can be probed in the most optimistic case for dimension-10 (dimension-8) operators can reach $\sim 10~(200)$ TeV at the upper limit of strongly coupled scenarios. For such a dimension-10 operator coefficient value, the dimension-8 operator coefficient corresponding to a little hierarchy in the vev with a 10 TeV EFT cut-off would then be expected to be $|\bar{c}_8^{(i)}| \sim 0.1$, which according to Fig.~\ref{fig:projections} is also within experimental reach.

Finally, we note that a vev contribution can also arise from dimension-8 operators contributing to dimension-6 positivity bounds~\cite{Remmen:2020uze, Remmen:2022orj, Davighi:2021osh}, which require additional UV assumptions but are experimentally more accessible. We leave a detailed phenomenological study of experimental prospects to future work.

\section{Conclusion}

\begin{figure}[t!]
    \centering
    \includegraphics[width=0.8\linewidth]{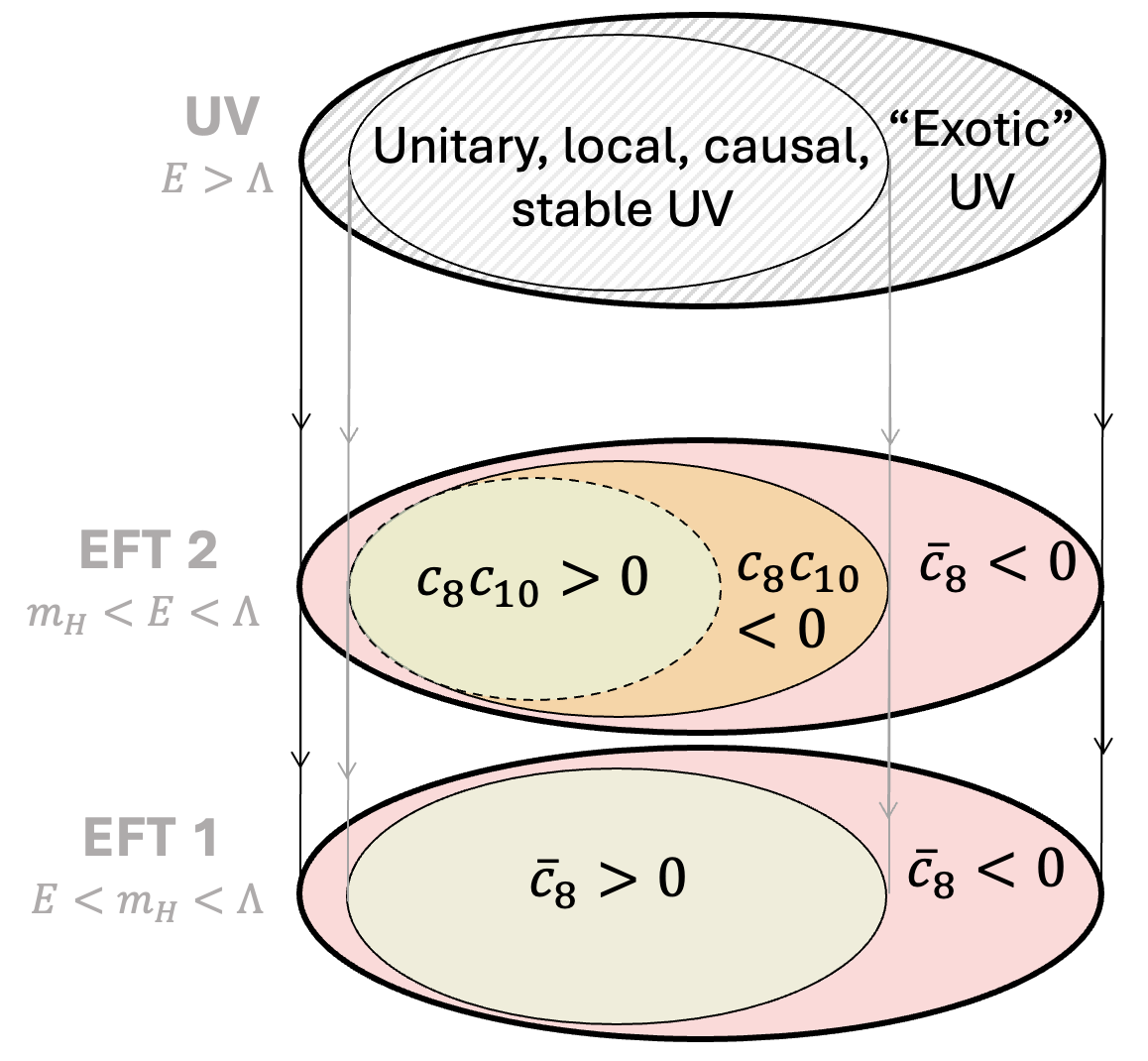}
    \caption{Illustration of how the assumptions of a unitary, local, and causal UV with a stable vacuum $v$ translate to constraints on the EFT parameter space in the IR. Region 1 where positivity is satisfied corresponds to the green subset in the lowest energy EFT 1 below a scalar mass $m_H$, which is split into further subsets in an EFT 2 at an intermediate scale above $m_H$ but below the EFT cut-off $\Lambda$. Scalar vev contributions are relevant for interpreting the bounds in the orange Region $\bar{1}$.  }
    \label{fig:UVEFTmap}
\end{figure}

\noindent We proposed a novel interpretation of positivity bounds when scalar vevs are taken into account. A positivity bound on a single higher-dimensional operator coefficient at low energies may subsume contributions from the vevs of scalars that only become apparent at higher energies in the next layer of EFT, where the scalar degrees of freedom can be produced on-shell. One consequence is the existence of a region of EFT parameter space where positivity is conditional upon a hierarchy in the scalar vev and the EFT cut-off. This is illustrated in Fig.~\ref{fig:UVEFTmap}.

While EFTs in this region feature an unnatural ratio of higher-dimensional operator coefficients, it is intriguing that such ratios may be related to fine-tuning in the electroweak hierarchy assuming only unitarity, causality and locality in the UV. There are not many other cases where potential phenomena in the IR, together with reasonable assumptions on the UV, lead to a restricted spectrum of allowed Higgs vevs. Perhaps the closest example of a UV assumption relating the Higgs vev to a different IR observable is the connection between a feeble fifth force and an upper bound on the Higgs vev assuming the weak gravity conjecture holds in the UV~\cite{Cheung:2014ega, Lust:2017wrl, Craig:2019fdy}. More generally, there can be other UV mechanisms that restrict the range of scalar field vevs, for instance the Swampland Distance Conjecture~\cite{Ooguri:2006in}. It is worthwhile exploring such unconventional relations between vevs in the IR and properties of the underlying UV physics that may help us better understand the hierarchy problem.

The possibility of living in the special EFT region we have identified may not be so far-fetched. After all, the coefficients of operators at dimensions 0, 2, and 4 in the Higgs potential all indicate that our universe is highly non-generic in many ways: not only is the Higgs quadratic term finely tuned to lie at the boundary of broken and unbroken phases, the Higgs quartic in the SM places us in a sliver of parameter space between vacuum stability and instability, while the cosmological constant value is precariously balanced between implosion and explosion. Dimension-6 operators could furthermore extend the connection between near-criticality and parameters of the SM~\cite{Giudice:2021viw, Khoury:2021zao, Steingasser:2023ugv}. It may well be that dimension-8 and -10 operators similarly place us on another boundary --- at the edge of positive and non-positive theory space.

\bigskip
\noindent {\it Acknowledgements.} --- TY was supported by a Branco Weiss Society in Science Fellowship and United Kingdom STFC grant ST/T000759/1. 
SM is supported by a UKRI Stephen Hawking Fellowship (EP/T017481/1). JD is funded by the European Research Council (ERC) under the European Union’s Horizon 2020 research and innovation programme under grant agreement 833280 (FLAY), and by the Swiss National Science Foundation (SNF) under contract 200020-204428. KM was supported by STFC grant
ST/T000759/1.


\end{document}